\documentclass[runningheads]{llncs}
\usepackage{graphicx}
\graphicspath{{figures/}}
\usepackage{subcaption}
\usepackage{xcolor}
\usepackage{booktabs}
\usepackage{tabularx}
\usepackage{multirow}

\usepackage[hidelinks]{hyperref}
\PassOptionsToPackage{hyphens}{url}
\usepackage{ifthen}
\usepackage{amssymb}
\newboolean{showcomments}
\setboolean{showcomments}{false} 
\ifthenelse{\boolean{showcomments}}
  {\newcommand{\nb}[2]{
    \fcolorbox{gray}{yellow}{\bfseries\sffamily\scriptsize#1}
    {\sf\small$\blacktriangleright$\textit{#2}$\blacktriangleleft$}
   } 
   
  }
  {\newcommand{\nb}[2]{}
   
  }

\begin{document}
\title{A Model-based Chatbot Generation Approach to Converse with Open Data Sources\thanks{Work supported by the Spanish government (TIN2016-75944-R project)}}
\titlerunning{A Model-based Chatbot Generation Approach for Open Data}

\author{Hamza Ed-douibi\inst{1}
\and Javier Luis Cánovas Izquierdo\inst{1} 
\and 
\\ Gwendal Daniel\inst{1} 
\and Jordi Cabot\inst{1,2}
}

\authorrunning{Ed-douibi et al.}

\institute{
UOC. Barcelona, Spain\\ \email{{\{hed-douibi,jcanovasi,gdaniel\}@uoc.edu}}\\ \and
ICREA. Barcelona, Spain \\ \email{jordi.cabot@icrea.cat}}
\maketitle              
\begin{abstract}
The Open Data movement promotes the free distribution of data. 
More and more companies and governmental organizations are making their data available online following the Open Data philosophy,
resulting in a growing market of technologies and services to help publish and consume data.
One of the emergent ways to publish such data is via Web APIs, which offer a powerful means to reuse this data and integrate it with other services.
\textsc{Socrata}, \textsc{CKAN} or \textsc{OData} are examples of popular specifications for publishing data via Web APIs.
Nevertheless, querying and integrating these Web APIs is time-consuming and requires technical skills that limit the benefits of Open Data movement for the regular citizen. 
In other contexts, chatbot applications are being increasingly adopted as a direct communication channel between companies and end-users.
We believe the same could be true for Open Data as a way to bridge the gap between citizens and Open Data sources.
This paper describes an approach to automatically derive full-fledged chatbots from API-based Open Data sources. 
Our process relies on a model-based intermediate representation (via UML class diagrams and profiles) to facilitate the customization of the chatbot to be generated.
\end{abstract}

\newlength{\mysep}
\setlength{\mysep}{0.35em}

\section{Introduction}
\label{sec:introduction}

Open Data has emerged as a movement that promotes the free distribution of data for everyone to consume and republish.
Governmental organizations are one of the significant sources of Open Data resources. They make their data publicly available online to provide more transparency and enable the general public to monitor and control the action of government bodies.
For instance, the Spanish Open Data portal\footnote{\url{https://datos.gob.es/en}} registers more than 20,000 resources while the European portal\footnote{\label{euurl}\url{https://www.europeandataportal.eu/data}}, which harvests the metadata of Public Sector Information available on public data portals across European countries, links to over 400,000.

On the one hand, Open Data promotes public awareness and aims at boosting citizen participation but, still, regular citizens hardly benefit from them as consuming Open Data requires non-trivial technical skills.
Indeed, more and more Open Data sources are released as ``web-friendly'' artifacts (e.g., LinkedData, APIs or NoSQL databases) that facilitate their consumption by external software applications and not directly by end-users.
In particular, some specific technologies to support the publication of Open Data in the Web have been widely adopted in the last years, namely: \textsc{Socrata}\footnote{\url{https://dev.socrata.com/}}, \textsc{CKAN}\footnote{\url{https://ckan.org/}} and \textsc{OData}\footnote{\url{https://www.odata.org/}}. 
Other organizations also rely on \textsc{OpenAPI}\footnote{\url{https://www.openapis.org/}}, an initiative to formally describe general-purpose REST APIs, to document their Open Data APIs.
While all these Web APIs ``standards'' offer a powerful means for writing advanced data queries, they require advanced technical knowledge that hampers their actual use by non-technical people.

On the other hand, chatbots are conversational agents typically embedded in instant messaging platforms. 
Users can ask questions or send requests to the chatbot using natural language, no need to learn any technical knowledge/language.
Chatbots have proven useful in various contexts to automate tasks and improve the user experience, such as automated customer services~\cite{chatbot-customer-service-2017} or education~\cite{chatbot-education-2007}.
Thus, we believe chatbots are the ideal interface to access and query Open Data sources, thus allowing citizens to access the government/company data they need directly. Citizens would ask the questions in their own language, and the chatbot would be the one in charge of translating that question into the corresponding API request/s. 

In this paper, we propose a model-based approach to generate chatbots tailored to the Open Data API technologies mentioned above.
The API definition is analyzed and imported as a UML schema annotated with UML profiles, which address specific domain information for chatbot configuration and Web API query generation.
This API model is then used to generate the corresponding chatbot to access and query the Open Data source.
To validate our approach, we provide a proof-of-concept Eclipse plugin that fully supports \textsc{Socrata} and allows the integration of other Open Data specifications (i.e., \textsc{OData}, \textsc{CKAN}) as well as generic web APIs (via \textsc{OpenAPI} specification).

The rest of the paper is organized as follows.
Section~\ref{sec:background} introduces the background of our work.
Section~\ref{sec:overview} briefly describes our approach while sections \ref{sec:importing} and \ref{sec:generator} describe its main phases, namely, Open Data Import and Bot Generation, respectively.
Section~\ref{sec:toolSupport} comments on the implemented tool support and Section~\ref{sec:related} presents the related work.
Finally, Section \ref{sec:conclusion} ends the paper and presents the future work. 

\section{Background}
\label{sec:background}

\subsection{Open Data}
The Open Data movement aims to make data free to use, reuse, and redistribute by anyone. 
In the last years, Open Data portals have evolved from offering data in text formats only (e.g., CSV, XML) towards web-based formats, such as LinkedData~\cite{bizer2011linked} and Web APIs, that facilitate the reuse and integration of Open Data sources by external web applications. 
In this subsection, we briefly describe the most common Web API technologies for Open Data, based on their popularity in governmental Open Data portals.

\vspace{\mysep}
\noindent \textsc{Socrata}. Promoted by Tyler Technologies, the \textsc{Socrata} data platform provides an integrated solution to create and publish Open Data catalogs.
\textsc{Socrata} supports predefined web-based visualizations of the data, the exporting of datasets in text formats and data queries via its own API that provides rich query functionalities through a SQL-like language called \textsc{SoQL}.
\textsc{Socrata} has been adopted by several governments around the world (e.g., Chicago\footnote{\url{https://data.cityofchicago.org}} or Catalonia\footnote{\url{http://governobert.gencat.cat/es/dades\_obertes/index.html}}). 

\vspace{\mysep}
\noindent \textsc{CKAN}. Created by the Open Knowledge Foundation, \textsc{CKAN} is an Open Source solution for creating Open Data portals and publishing datasets in them. 
As an example, the European Data Portal relies on CKAN.
Similar to \textsc{Socrata}, CKAN allows viewing the data in Web pages, downloading it, and querying it using a Web API.
The CKAN DataStore API can be used for reading, searching, and filtering data in a classical Web style using query parameters or by writing SQL statements directly in the URL.

\vspace{\mysep}
\noindent \textsc{OData}. Initially created by Microsoft, \textsc{OData} is a protocol for creating data-oriented REST APIs with query and update capabilities.
\textsc{OData} is now also an OASIS standard.
It is especially adapted to expose and access information from a variety of data sources such as relational databases, file systems, and content management systems.
\textsc{OData} allows creating resources that are defined according to a data model and can be queried by Web clients using a URL-based query language in a SQL-like style.
Many service providers adopted and integrated \textsc{OData} in their solutions (e.g., \textsc{SAP} or \textsc{IBM WebSphere}).

\vspace{\mysep}
\noindent \textsc{OpenAPI}. Evolving from Swagger, the \textsc{OpenAPI} specification has become the \emph{de facto} standard to describe REST APIs.
Though not specific for Open Data, \textsc{OpenAPI} is commonly used to specify all kinds of Web APIs, including Open Data ones (e.g., Deutsche Bahn\footnote{\url{https://developer.deutschebahn.com/store}}). 

\vspace{\mysep}
In our approach, we target Open Data Web APIs described by any of the previous solutions.
We rely on model-driven techniques to cope with the variety of data schema and operation representations, as described in the next sections. 

\subsection{Chatbots}
Chatbots are conversational interfaces able to employ Natural Language Processing (NLP) techniques to ``understand'' user requests and reply accordingly, either by providing a textual answer and/or executing additional external/internal services as part of the fulfillment of the request.

NLP covers a broad range of techniques that may combine parsing, pattern matching strategies and/or Machine Learning (ML) to represent the chatbot knowledge base. 
The latter is the dominant one at the moment thanks to the popularization of libraries and Cloud-based services like \textsc{DialogFlow}\footnote{\url{https://dialogflow.com}} or \textsc{IBM Watson Assistant}\footnote{\url{https://www.ibm.com/cloud/watson-assistant}}, which rely on neural networks to match user intents.

However, chatbot applications are much more than raw language processing components~\cite{chatbot-lessons-2018}. 
Indeed, the conversational component of the application is usually the front-end of a larger system that involves data storage and service integration and execution as part of the chatbot reaction to the user intent.
Thus, we define a chatbot as an application embedding a \emph{recognition engine} to extract \emph{intentions} from user inputs, and an \emph{execution component} performing complex event processing represented as a set of \emph{actions}. 

\emph{Intentions} are named entities that can be matched by the recognition engine. 
They are defined through a set of \emph{training sentences}, which are input examples used by the recognition engine's ML/NLP framework to derive a number of potential ways the user could use to express the intention\footnote{In this article we focus on ML/NLP-based chatbots, but the approach can be applied to alternative recognition techniques.}. 
Matched intentions usually carry \emph{contextual information} computed by additional extraction rules (e.g. a typed attribute such as a city name, a date, etc) available to the underlying application. 
In our approach, \emph{Actions} are used to represent simple responses such as sending a message back to the user, as well as advanced features required by complex chatbots like database querying or external service calling (e.g. API queries in this paper).
Finally, we define a \emph{conversation path} as a particular sequence of received user \emph{intentions} and associated \emph{actions} (including non-messaging actions) that can be executed by the chatbot application.

\section{Overview}
\label{sec:overview}

\begin{figure}[t]
  \centering
  \includegraphics{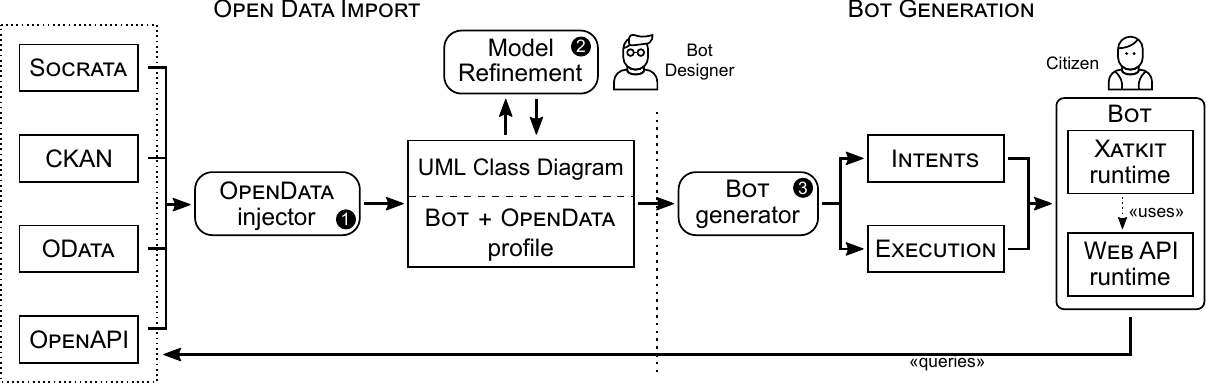}
  \caption{Overview of our approach.}
  \label{fig:overview}
\end{figure}

In this section, we present an overview of our proposal, depicted in Figure~\ref{fig:overview}.
Our proposal is split into two main phases, \emph{Open Data Import} and \emph{Bot Generation}.

During the import phase, an Open Data API model is injected (see \emph{\textsc{OpenData} injector}) and refined (see \emph{Model refinement}). 
The injector supports several input formats (i.e., \textsc{Socrata}, \textsc{CKAN}, \textsc{OData} and \textsc{OpenAPI}) and the result is a unified model representation of the API information (i.e., operations, parameters and data schemas). 

Without loss of generality, this inferred API model is expressed as a UML class diagram to represent the API information plus two additional UML profiles\footnote{Profiles are a lightweight extension mechanism to add additional semantics to UML models, part of the UML standard}. 
The first one, the \textsc{Open Data} profile, is used to keep track of technical information on the input source (e.g., to be used later on for the Bot to know which API endpoint to call and how). 
The second one is the \textsc{Bot} profile, proposed to annotate the model with bot-specific configuration options 
allowing for a more flexible chatbot generation.
Once the injector finishes, the \emph{Bot Designer} refines the obtained model using this second profile.
During this step, elements of the API can be hidden, their type can be tuned, or synonyms can be provided (so that the chatbot knows better how to match requests to the to data elements).\looseness=-1

The generation phase is in charge of creating the chatbot definition (see \emph{Bot Generation}). 
This phase involves specifying both the bot intentions and its response actions. 
In our scenario, responses involve calling the right Open Data API operation/s, processing the answer, and presenting it back to the user. 

As bot platform we use \textsc{Xatkit}~\cite{xatkit}, a flexible multi-platform (chat)bot development framework, though our proposal is generic enough to be adapted to work with other chatbot frameworks.
\textsc{Xatkit} comprises two main Domain-Specific Languages (DSLs) to define bots: 
\textsc{Intent DSL}, which defines the user inputs through training sentences, and context parameter extraction rule (see \emph{Intents});
and \textsc{Execution DSL}, in charge of expressing how the bot should respond to the matched intents (see \emph{Execution}).

\textsc{Xatkit} comes with a runtime to interpret and execute the bots' definitions. 
The execution engine includes several connectors to interact with external platforms (e.g., \textsc{Slack} or \textsc{Github}). 
In the context of this work, we implemented a new runtime in \textsc{Xatkit} to enable the communication with Web APIs.

The next sections describe each of these components in more detail. 
We will use the following running example to illustrate them.
The example is based on an API provided by the Transparency Portal of Catalonia.
In particular, the API that gives access to pollution data gathered by the surveillance network deployed within Catalonia.
The data registers the air quality in Catalonia from 1991 until now, and it is updated daily. 
Besides the concentration of pollutants in the air, it is also possible to query the location and type of the measurement stations.
The API has been specified following the \textsc{Socrata} v2.1 specification\footnote{\url{https://dev.socrata.com/foundry/analisi.transparenciacatalunya.cat/uy6k-2s8r}}.

\section{Importing Open Data APIs as Models}
\label{sec:importing}

The import phase starts by analyzing the Open Data API description to inject a UML model representing its concepts, properties, and operations. This model is later refined by the bot designer.
Next sections describe the main elements of this process.
We will introduce first the modeling support required to represent Open Data APIs, then the injection step and finally the main tasks to tackle in the refinement step.

\subsection{Modeling Open Data APIs}
To model Open Data APIs, we propose employing UML class diagrams plus two UML profiles required to optimize and customize the bot generation. 

\subsubsection{Core Open Data representation as a UML Class Diagram}
\begin{figure}[t]
  \centering
  \includegraphics{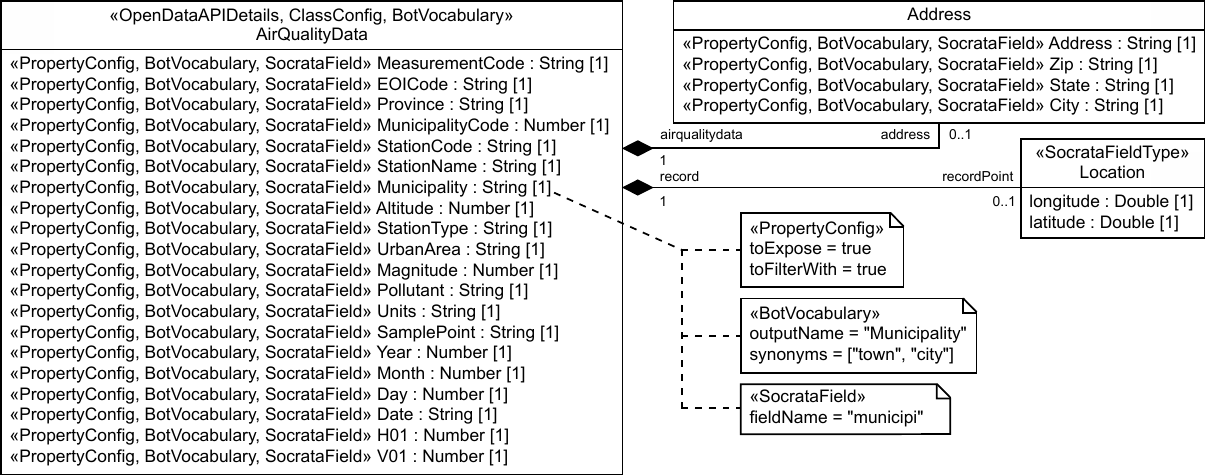}
  \caption{UML model for the running example (our editor can show/hide the stereotypes to show a simplified representation of the diagram).}
  \label{fig:example}
  \vspace{0.5em}
\end{figure}

Concepts, properties and operations of Open Data APIs are represented using standard elements of UML structural models (classes, properties and operations, respectively). 
Figure \ref{fig:example} shows an excerpt of the UML model for the running example\footnote{Full model at \url{http://hdl.handle.net/20.500.12004/1/C/ER/2020/575}}. 
As can be seen, the model includes the core concept of the API, called \emph{AirQualityData}; plus two more classes to represent data structures (i.e., \emph{Address} and \emph{Location}).
Note that the some elements include the stereotypes that we will present later. 

It is worth noting that most Open Data APIs focus around a single core data element composed of a rich set of properties which can be split (i.e., ``normalized'') into separate UML classes following good design practices, also facilitating the understanding of the model.
This is what we have done for the UML diagram shown in Figure \ref{fig:example}.

\subsubsection{The \textsc{Bot} profile}
To be able to generate more complete bots, in particular, to expand on aspects important for the quality of the conversation, the \textsc{Bot} profile adds a set of stereotypes for UML model elements that cover 
(1) what data should the chatbot expose, 
(2) how to refer to model elements (instead of the some obscure internal API identifiers), 
and (3) synonyms for model elements that citizens may employ when attempting to alternatively name the concept as part of a sentence.

\begin{figure}[t]
  \centering
  \includegraphics{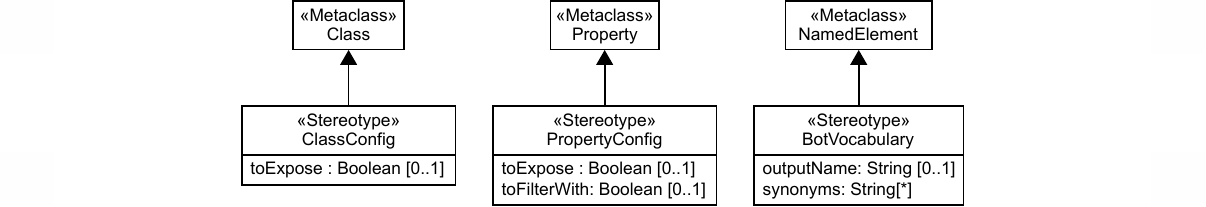}
  \caption{\textsc{Bot} profile.}
  \label{fig:botProfile}
\end{figure}

Figure~\ref{fig:botProfile} shows the specification of the \textsc{Bot} profile. 
It comprises three stereotypes, namely, \emph{ClassConfig}, \emph{PropertyConfig} and \emph{BotVocabulary}, extending the \emph{Class}, \emph{Property} and \emph{NamedElement} UML metaclasses, respectively.
The \emph{ClassConfig} stereotype includes the \emph{toExpose} property, in charge of defining if the annotated Class element has to be made visible to end-users via the chatbot.
The \emph{PropertyConfig} stereotype also includes the \emph{toExpose} property, with the same purpose; plus the \emph{toFilterWith} property, which indicates if the corresponding annotated property can be used to filter results as part of a conversation iteration.
For instance, in our running example, pollution data could be filtered via date.
Finally, the \emph{BotVocabulary} stereotype can annotate almost any UML model element and allows specifying a more ``readable'' name to be used when printing concept information and a set of synonyms for the element. \looseness=-1

In Figure~\ref{fig:example} we see the \textsc{Bot} profile applied on the running example. Note, for instance, how we define that \emph{town} and \emph{city} could be used as synonyms of \emph{Municipality} and that this attribute can be used to filter \emph{AirQuality} results.


\subsubsection{The \textsc{OpenData} profile}
While the previous profile is more oriented towards improving the communication between the chatbot and the user, this \textsc{OpenData} profile is specially aimed at defining the technical details the chatbot needs to know in order to communicate with the Open Data API backend.
The profile defines a set of stereotypes that cover how to access the information of the model elements via the Web API. 
The access method depends on the specification followed by the Open Data API, which can be \textsc{Socrata}, \textsc{CKAN}, \textsc{OData} or \textsc{OpenAPI}.

\begin{figure}[t]
  \centering
  \includegraphics{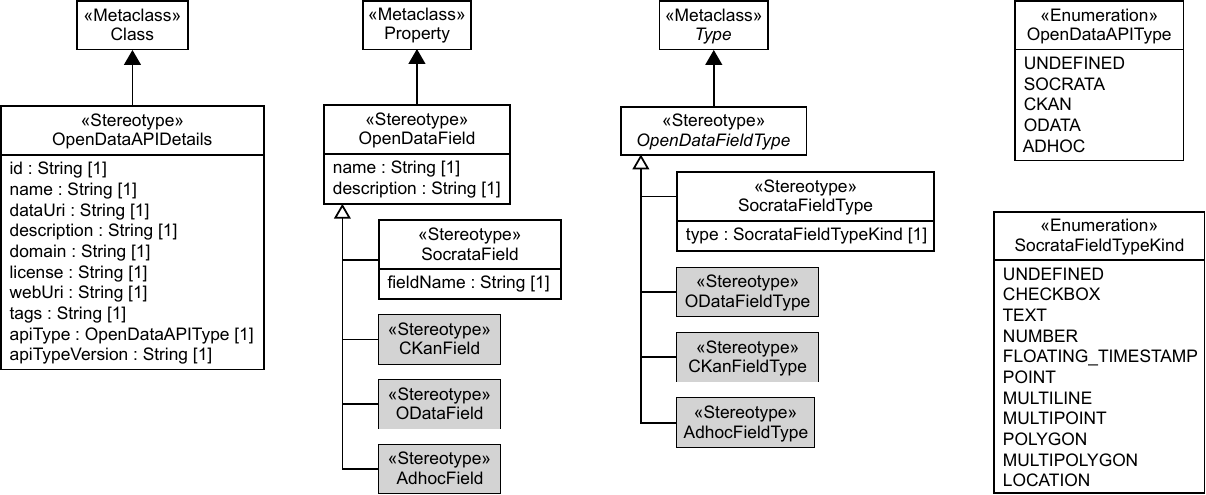}
  \caption{\textsc{OpenData} profile.}
  \label{fig:opendataProfile}
\end{figure}

Figure~\ref{fig:opendataProfile} shows the \textsc{OpenData} profile.
As can be seen, we have defined three stereotypes, namely, \emph{OpenDataAPIDetails}, \emph{OpenDataField} and \emph{OpenDataFieldType}, which extend \emph{Class}, \emph{Property} and \emph{Type} UML metaclasses, respectively.
The \emph{OpenDataAPIDetails} stereotype includes a set of properties to enable the API query of the annotated UML Class.
For instance, it includes the \emph{domain} and \emph{webUri} to specify the host and route parameters to build the query.
It also includes the \emph{APIType} property, which sets the kind of Open Data API (see values of the \emph{OpenDataAPIType} enumeration).
The \emph{OpenDataField} stereotype annotates properties with additional information depending on the type of Open Data API used. 
For instance, the \emph{SocrataField} stereotype indicates the name of the field (see \emph{fieldName}) that has to be queried to retrieve the annotated property. 
Finally, the \emph{OpenDataFieldType} stereotype includes additional information regarding the types of the properties used by the Open Data APIs. 

Figure~\ref{fig:opendataProfile} also includes stereotypes prefixed with \emph{CKAN}, \emph{OData} and \emph{Adhoc} (in grey) to cover the information required for \textsc{CKAN}, \textsc{OData} and \textsc{OpenAPI} specifications.
We do not fully detail them due to the lack of space but they are available online\footnote{\url{http://hdl.handle.net/20.500.12004/1/C/ER/2020/822}}.  
Besides, the \emph{Adhoc} annotations also use the \textsc{OpenAPI} profile~\cite{DBLP:conf/models/Ed-DouibiIBC19}.\looseness=-1

As an example, this profile is also used to annotate Figure~\ref{fig:example}. While the profile is rather exhaustive and comprises plenty of detailed, technical information, note that it is automatically applied during the injection process. 

\subsection{Injection of Open Data Models}
Injectors collect specific data items from the API descriptions in order to generate a model representation of the API.
In a nutshell, regardless of the API specification used, the injector always collects information about the API metadata, its concepts and properties.
This information is used to generate a UML model annotated with the \textsc{OpenData} profile. 
Additionally, injectors also initialize the annotations corresponding to the \textsc{Bot} profile with default values 
which will later be tuned during the refinement step (see next subsection). 

In our running example, the injector takes as input the \textsc{Socrata} description of the data source\footnote{\url{https://analisi.transparenciacatalunya.cat/api/views/metadata/v1/uy6k-2s8r.json}} to create the UML model classes and stereotypes.
To complement the definition of the data fields and their types, the injector also calls the \textsc{Views API}\footnote{\url{https://analisi.transparenciacatalunya.cat/api/views.json?id=uy6k-2s8r}}, an API provided by \textsc{Socrata} to retrieve metainformation about the data fields of datasets.


\subsection{Refinement of Open Data Models}
Once the injection process creates a UML schema annotated with stereotypes, the bot designer can revise and complete it to generate a more effective chatbot.
The main refinement tasks cover: 
(a) providing default names and synonyms for model elements, which enriches the way the chatbot (and the user) can refer to such elements;
and (b) set the visibility of data elements, thus enabling the designer to hide some elements of the API in the conversation.

During the refinement step, the bot designer can also revise the \textsc{OpenData} profile values if the API description is not fully aligned with the actual API behavior, as sometimes the specification (input of the process) unfortunately is not completely up-to-date with the API implementation deployed (e.g., type mismatchings).

\section{Generating the Bot}
\label{sec:generator}

The generation process takes the annotated model as input and derives the corresponding chatbot implementation.
As our proposal relies on \textsc{Xatkit}, this process generates the main artifacts required by such platform, specifically: 
(1) \emph{intents} definition, which describes the user intentions using training sentences (e.g., the intention to retrieve a specific data point from the data source, or to filter the results), contextual information extraction, and matching conditions;
and (2) \emph{execution} definition, which specifies the chatbot behavior as a set of bindings between user intentions and response actions (e.g., displaying a message to answer a question, or calling an API endpoint to retrieve data). A similar approach could be followed when targeting other chatbot platforms as they all require similar types of input artefact definitions in order to run bots.

The main challenge when generating the chatbot implementation is to provide effective support to drive the conversation. 
To this aim, it is crucial to identify both the topic/s of the conversation and the aim of the chatbot, which will enable the definition of the conversation path.
In our scenario, the topic/s is set by the API domain model (i.e., the vocabulary information embedded in the UML model and the \textsc{Bot} profile annotations) while the aim is to query the API endpoints (relying on the information provided by the \textsc{OpenData} profile).

Our approach supports two conversation modes, which are implemented in the \emph{intents} file. 
Table~\ref{tab:intents} lists the main intents generated for the conversation, which we will present while describing the conversation modes.

\begin{figure}[t]
\begin{subfigure}{.5\textwidth}
  \centering
  \includegraphics{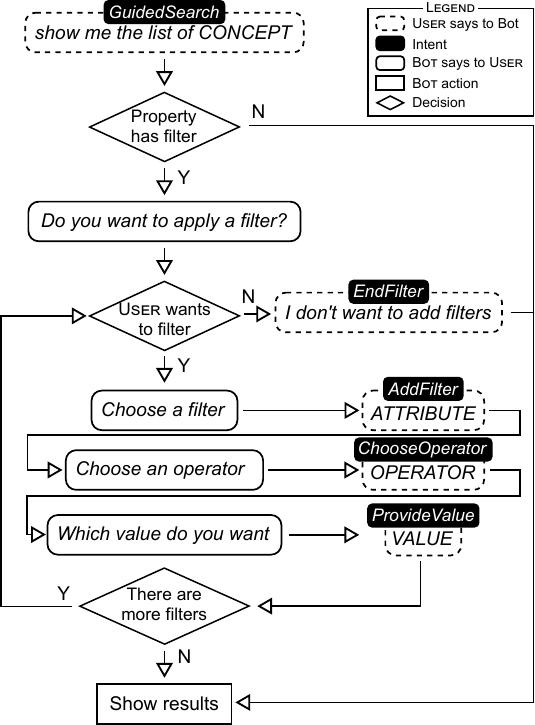}  
  \caption{}
  \label{fig:flow:diagram}
\end{subfigure}
\begin{subfigure}{.2\textwidth}
  \includegraphics[width=\textwidth]{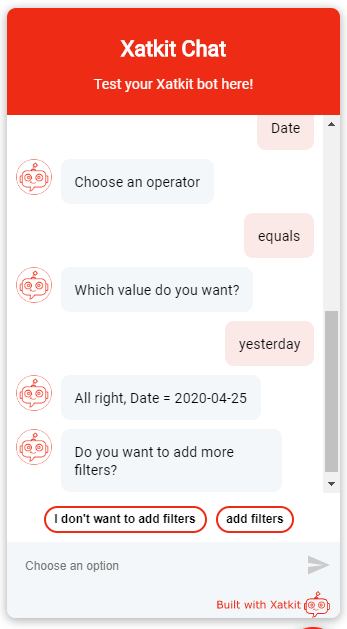}  
  \caption{}
  \label{fig:flow:example1}
\end{subfigure}
\begin{subfigure}{.2\textwidth}
  \includegraphics[width=\textwidth]{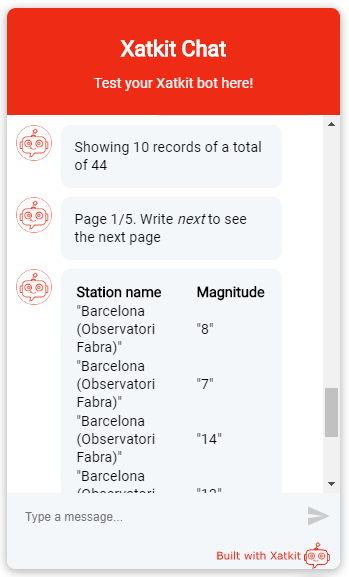}  
  \caption{}
  \label{fig:flow:example2}
\end{subfigure}
\caption{(a) Conversation path in guided queries. (b) A guided conversation. (c) Showing the results.}
\label{fig:flow}
\end{figure}

\begin{description}
\item[Direct queries]
The most basic communication in a chatbot is when the user directly asks what is needed (e.g., \emph{What was the pollution yesterday?}).
To support this kind of query, we generate intents for each class and attribute in the model\footnote{Note that this scales well as we do not actually create completely separate intents for each possible combination but use intent templates that can be instantiated at run-time over the list of elements in the model.} enabling users to ask for that specific information. Moreover, we also generate filtering intents that help users choose a certain property as filter to cope with queries returning too many data. Table~\ref{tab:intents}, row 1, shows an example of this type of direct intent generated and a possible user utterance (i.e., concrete user input query) corresponding to this intent kind.



\item[Guided queries]
We call \emph{guided queries} those interactions where there is an exchange of questions/requests between the chatbot and the user, simulating a more natural Open Data exploration approach. 
Their implementation require a clear definition of the possible dialog paths driving the conversation.
Table~\ref{tab:intents}, rows 2-6, shows the intents generated for guided conversations, which are applied in order (starting with \emph{GuidedSearch} and then adding filters using the rest of the intents).   
Figure~\ref{fig:flow:diagram} aims to summarize the possible conversation paths and the application order of the intents. 
The shown paths start once the user asks for a specific concept made available by the API. 
If the property can be filtered, the path gives the user the option to apply such a filter. 
This step repeats while there are other filtering options. 
Figure~\ref{fig:flow:example1} shows an example of guided query for our running example.
\end{description}


\begin{table*}[t]
\caption{Main intents generated.}
\label{tab:intents}
\fontsize{7pt}{8pt}\selectfont
\begin{tabularx}{\textwidth}{c@{\hspace{0.5em}} c@{\hspace{0.5em}} l@{\hspace{0.5em}} l}%
\multicolumn{1}{c}{\textsc{Mode}} & 
\multicolumn{1}{c}{\textsc{Intent}} & 
\multicolumn{1}{c}{\textsc{Description}} & 
\multicolumn{1}{c}{\textsc{Example Sentence}} \\
\toprule
\multirow{2}{*}{Direct} & \multirow{2}{*}{\emph{DirectSearch}}                   & \multirow{2}{*}{Shows elements given a filter}                   & \texttt{show me all the air quality data with} \\
                        &                                                        &                                                                  & \texttt{municipality equals to "Barcelona"} \\
\midrule
Guided                  & \emph{GuidedSearch}                                    & Shows elements in conversation                                   & \texttt{show me the list of air quality data} \\
\midrule
Guided                  & \emph{AddFilter}                                       & Chooses an attribute to filter                                   & \texttt{date} \\
\midrule
Guided                  & \emph{ChooseOperator}                                  & Chooses an operator                                              & \texttt{equals}  \\
\midrule
Guided                  & \emph{ProvideValue}                                    & Sets a value                                                     & \texttt{yesterday} \\
\midrule
Guided                  & \emph{EndFilter}                                       & Ends filter for results                                          & \texttt{I don't want to add filters} \\
\midrule
Both                    & \emph{SelectField}                                     & Select fields for results                                        & \texttt{municipality} \\
\midrule
Both                    & \emph{ShowResult}                                      & Ends field selection for results                                 & \texttt{I don't want to add fields} \\
\midrule
Both                    & \emph{AddPostFilter}                                   & Adds a filter in results                                         & \texttt{add filter magnitude less than "14"} \\
\midrule
\multirow{2}{*}{Both}   & \multirow{2}{*}{\emph{SortOrderBy}} 	                 & \multirow{2}{*}{Sorts/Orders the result}                         & \texttt{sort by name ASC} \\
                        &                                     	                 &                                                                  & \texttt{order by date ASC} \\
\midrule
Both                    & \emph{NextPage}                     	                 & Shows the next page of results                                   & \texttt{show me next page}\\
\midrule
Both 					          & \emph{AddPostFunction}                                 & Calls function on results                                        & \texttt{calculate FUN ATT} \\
\bottomrule
\end{tabularx}
\vspace{-0.7cm}
\end{table*}

As input assistance, both direct and guided modes include buttons as shortcuts in the conversation interface (see Figure~\ref{fig:flow:example1}). Once the chatbot collects the request (with the possible filters) from the user, the next step is to query the involved Open Data Web API, which relies on the information provided in the \textsc{OpenData} profile.
The implementation of this step is specified in the \emph{execution} file, where the steps to query, filter and recover the information from the API are generated. 


The final step in every query performed by the chatbot involves presenting and post-processing the results.
In the presentation step, the user indicates the fields to show. 
Table~\ref{tab:intents}, rows 7-8, shows the intents for setting the fields to present.
Figure~\ref{fig:flow:example2} shows an example of the result for our running example showing the fields \emph{Municipality} and \emph{Magnitude}.
In the post-processing step, the user can apply additional filters, sort the results and paginate them.
Table~\ref{tab:intents}, rows 9-12, shows the intents for post-processing the results.
Finally, note that our approach also incorporates aggregation functions (e.g., calculate the average, minimum o maximum) as post-processing operators.

\section{Tool support}
\label{sec:toolSupport}

Our approach has been implemented as a new plugin for the Eclipse platform\footnote{\url{https://github.com/opendata-for-all/open-data-chatbot-generator}}.
We rely on the environment extensibility and modeling support provided by Eclipse to import and generate the chatbot definition, which is then eventually executed by \textsc{Xatkit}. 

\begin{figure}[t]
\begin{subfigure}{.5\textwidth}
  \centering
  \includegraphics[width=0.5\textwidth]{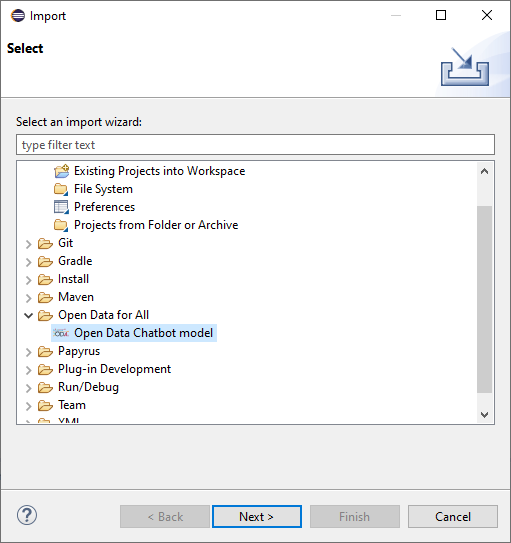}  
  \caption{}
  \label{fig:tool:ex1}
  \vspace{1.2em}
\end{subfigure}
\begin{subfigure}{.5\textwidth}
  \centering
  \includegraphics[width=0.5\textwidth]{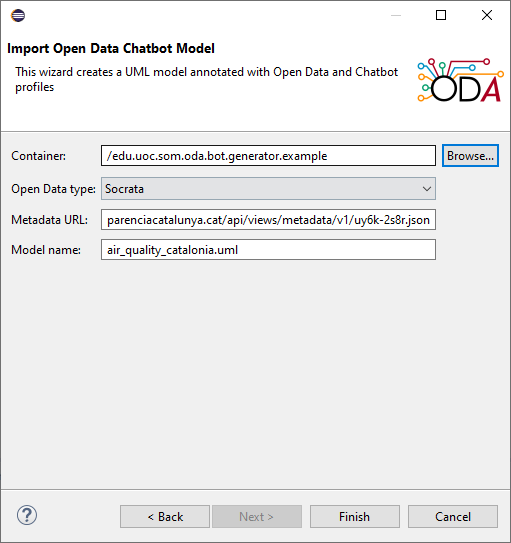}  
  \caption{}
  \label{fig:tool:ex2}
  \vspace{1.2em}
\end{subfigure}
\begin{subfigure}{.5\textwidth}
  \centering
  \includegraphics[width=\textwidth]{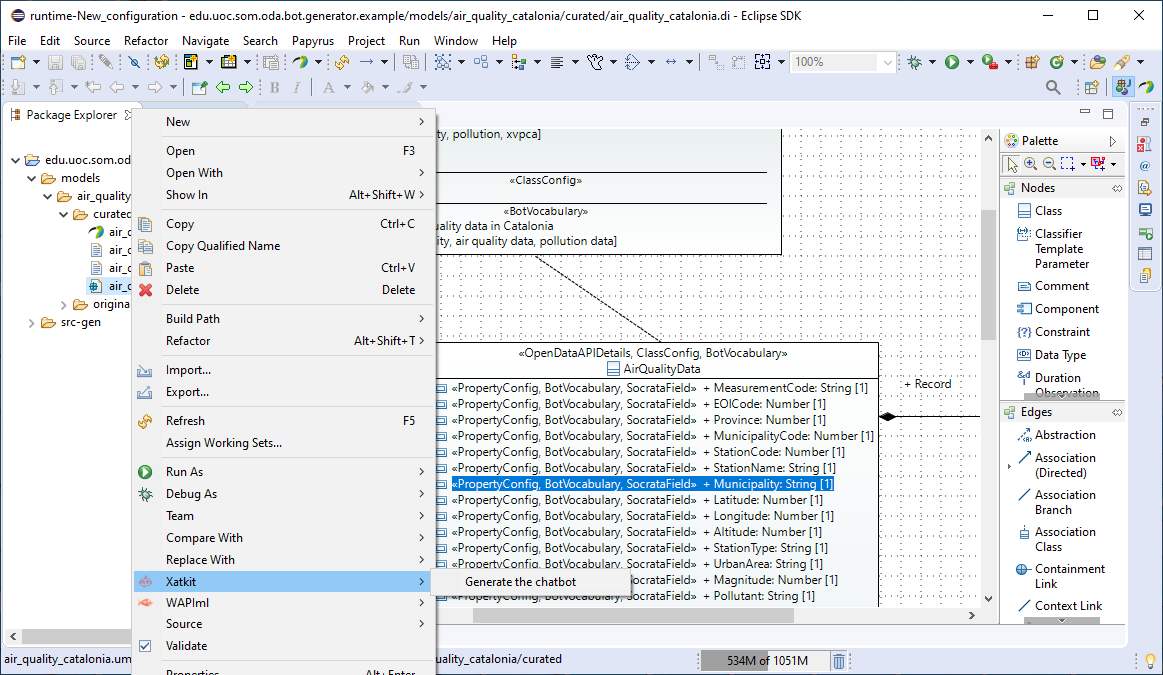}  
  \caption{}
  \label{fig:tool:ex3}
\end{subfigure}
\begin{subfigure}{.5\textwidth}
  \centering
  \includegraphics[width=\textwidth]{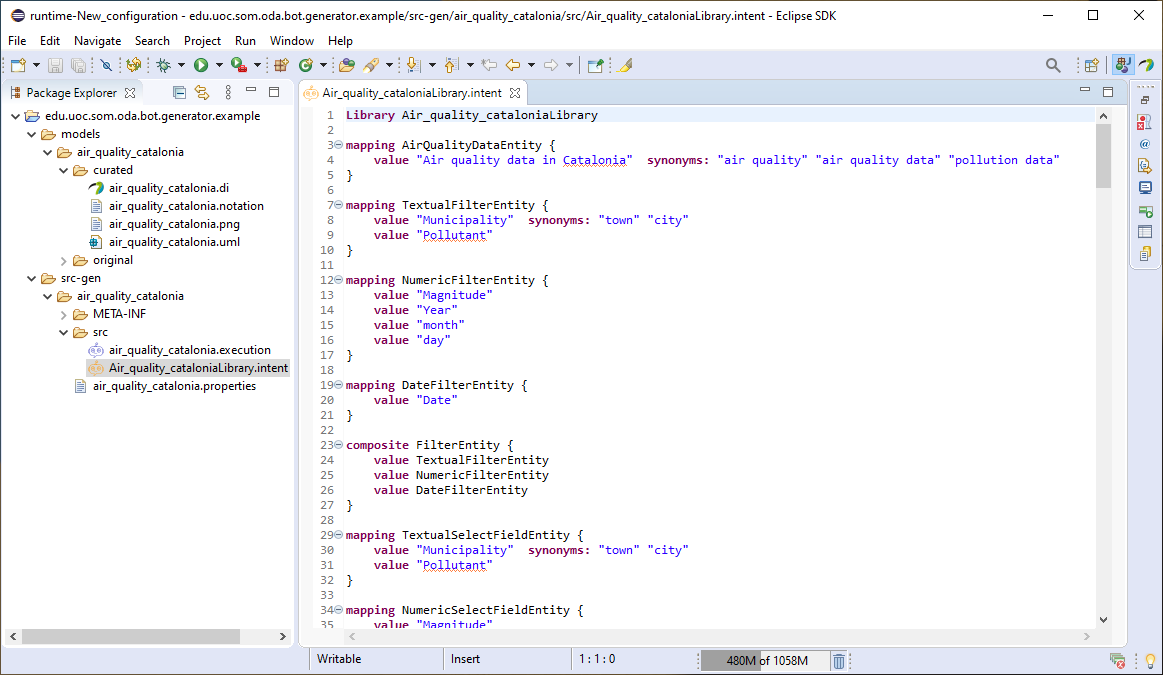}  
  \caption{}
  \label{fig:tool:ex4}
\end{subfigure}
\caption{Screenshots of the tool support: (a) and (b) import wizard, (c) generation wizard and (d) generated bot.}
\label{fig:tool}
\end{figure}

Figure~\ref{fig:tool} shows several screenshots of the development environment. 
It comprises two wizards to perform the import and generation phases.
During the import phase, the UML model is loaded (see wizard in Figures~\ref{fig:tool:ex1} and~\ref{fig:tool:ex2}) visualized and refined using the Papyrus modeling IDE\footnote{\url{https://www.eclipse.org/papyrus}}.
Once completed, our generation wizard (see Figures~\ref{fig:tool:ex3} and ~\ref{fig:tool:ex4}) creates the definition of the chatbot.
To enable \textsc{Xatkit} to run this new kind of open data bots, we have extended it with a new component\footnote{\url{https://github.com/xatkit-bot-platform/xatkit-rest-platform}}  to communicate with Web APIs.

\section{Related Work}
\label{sec:related}

The role of chatbots in Open Data has not been widely studied in the literature. 
Keyner \emph{et al.}~\cite{DBLP:conf/esws/KeynerSV19} proposed a chatbot to help users find data sources in an Open Data repository by relying on geo-entity annotations. 
However, the chatbot only suggests the data sources to explore. 
It does not provide querying capabilities to consult those data sources. 
The work by Neumaier \emph{et al.}~\cite{NeumaierSV17} is similar, also focusing on the suggestion of potential useful datasets. 
The work by Porreca \emph{et al.}~\cite{PorrecaLMVC17} describes a case study of using a chatbot for a concrete dataset. 
In all cases, chatbots are manually created.

A couple of works address the creation of chatbots to query Web APIs, especially \textsc{OpenAPI}-based ones. 
Our own \textsc{OpenAPI} bot~\cite{Ed-DouibiIBC19} helps developers understand what they could do with an API for which its \textsc{OpenAPI} definition is available. 
Instead, the work by Vazir \emph{et al.}~\cite{VaziriMSSH17} generates a chatbot to facilitate the execution of calls to the API itself. 
Nevertheless, they remain very implementation-oriented and do not offer any abstraction mechanism to add further semantics and flexibility to the bot generation process, as we do. 

Chatbot modeling and generation has also been proposed in some works (i.e., \cite{CastaldoDMZ19,Sindhgatta,EMMSAD20}) but none of these works proposes an end-to-end approach as ours, from the reverse engineering of the Open Data source to the generation of a chatbot actually able to directly call the initial source. 

Therefore, to the best of our knowledge, ours is the first work aimed at automatically generating chatbots to directly interact with Open Data sources using a model-based approach.


\section{Conclusion}
\label{sec:conclusion}

In this paper, we have presented a model-based approach to generate chatbots as user-friendly interfaces to query Open Data sources published as web APIs.
The resulting chatbot accepts both direct queries and guided conversations, where the chatbot and the user interact to precise the final query to send to the API.
We have implemented our approach as an Eclipse plugin that fully supports \textsc{Socrata} and allows the integration of other Open Data specifications (i.e., \textsc{OData}, \textsc{CKAN}) as well as generic web APIs (via \textsc{OpenAPI} specification); and generates chatbots using the \textsc{Xatkit} platform.

As further work, we plan to work on several extensions of this core framework:

\noindent \textbf{Support for advanced queries}. 
Our approach supports mainly descriptive queries where users navigate the data sources to learn about the facts explicitly stated there. 
However, there are other types of queries also interesting from an open data perspective; for instance, we could have:
(i) diagnostic queries, which focus on the analysis of potential reasons for a fact to happen;
(ii) predictive queries, aimed at exploring how a fact may evolve in the future;
and (iii) prescriptive ones, which study how to reproduce a fact.
We plan to define more advanced query templates to provide an initial support for these other types of queries.\looseness=-1

\vspace{\mysep}
\noindent \textbf{Composition of several Open Data sources}.
Many times, the data needs of a citizen span several Web APIs. 
The chatbot should be able to query and combine those different sources, dealing with potential composition links among them.
This composition is not trivial and involves the well-known challenges of any data integration scenario (e.g., entity matching) plus some others more API-specific like finding the optimal paths (even based on non-functional properties), as sometimes similar information can be obtained from different overlapping sources. Existing works on API composition~\cite{DBLP:conf/esocc/Ed-DouibiIC18,DBLP:conf/icws/MusyaffaHSOA16,DBLP:conf/esocc/CremaschiP17} can be used here.  

\vspace{\mysep}
\noindent \textbf{Voice-driven chatbots}. 
The growing adoption of smart assistants emphasizes the need to design chatbots supporting not only text-based conversations but also voice-based interactions. 
We believe that our chatbot could benefit from such a feature to improve the citizen's experience further when manipulating Open Data APIs. 
While \textsc{Xatkit}'s modular architecture supports both textual and voice-based chatbots, additional research is required to translate raw data returned by the API into sentences that can be read by the bot. 

\vspace{\mysep}
\noindent \textbf{Additional types of data sources}. 
We cover the most common choices in governmental Open Data portals, but they are not the only ones.
For instance, \textsc{LinkedData} sources, pure \textsc{RDF} files, \textsc{GeoJSON} collections, or database dumps, among others, are also used. 
We plan to develop additional import components that can target these technologies and integrate them into our framework.

\vspace{-0.6em}

\bibliographystyle{splncs04}
\bibliography{2020-ER}

\end{document}